\documentclass[reprint,aps,prb,showpacs,floatfix,superscriptaddress,a4paper,citeautoscript,longbibliography]{revtex4-1}

\usepackage{amsmath, amsfonts, amssymb, amsbsy}
\usepackage{graphicx}
\usepackage{xspace}

\usepackage{hyperref}

\hypersetup{colorlinks=true, linkcolor=blue, citecolor=blue,urlcolor=blue}

\newcommand{\SrIrOSL}{Sr$_{2}$IrO$_4$\xspace}
\newcommand{\SrIrOBL}{Sr$_{3}$Ir$_{2}$O$_7$\xspace}

\newcommand{\eg}{\emph{e.\,g.}\xspace}

\begin{document}
\title{The $J_{\mathrm{eff}}=\frac{1}{2}$ insulator \SrIrOBL studied by means of angle-resolved photoemission spectroscopy}

\author{B.~M. Wojek}
\homepage{http://bastian.wojek.de/}
\affiliation{KTH Royal Institute of Technology, ICT Materials Physics, Electrum 229, 164 40 Kista, Sweden}
\author{M.~H. Berntsen}
\affiliation{KTH Royal Institute of Technology, ICT Materials Physics, Electrum 229, 164 40 Kista, Sweden}
\author{S.~Boseggia}
\affiliation{London Centre for Nanotechnology and Department of Physics and Astronomy, University College London, London WC1E 6BT, United Kingdom}
\affiliation{Diamond Light Source Ltd., Oxfordshire OX11 0DE, United Kingdom}
\author{A.~T. Boothroyd}
\affiliation{Clarendon Laboratory, Department of Physics, University of Oxford, Parks Road, Oxford OX1 3PU, United Kingdom}
\author{D.~Prabhakaran}
\affiliation{Clarendon Laboratory, Department of Physics, University of Oxford, Parks Road, Oxford OX1 3PU, United Kingdom}
\author{D.~F. McMorrow}
\affiliation{London Centre for Nanotechnology and Department of Physics and Astronomy, University College London, London WC1E 6BT, United Kingdom}
\author{H.~M. R\o{}nnow}
\affiliation{Institute of Condensed Matter Physics (ICMP), \'Ecole Polytechnique F\'ed\'erale de Lausanne (EPFL), 1015 Lausanne, Switzerland}
\author{J.~Chang}
\affiliation{Institute of Condensed Matter Physics (ICMP), \'Ecole Polytechnique F\'ed\'erale de Lausanne (EPFL), 1015 Lausanne, Switzerland}
\affiliation{Swiss Light Source, Paul Scherrer Institut, 5232 Villigen, Switzerland}
\author{O.~Tjernberg}
\affiliation{KTH Royal Institute of Technology, ICT Materials Physics, Electrum 229, 164 40 Kista, Sweden}

\date{\today}

\begin{abstract}
The low-energy electronic structure of the $J_{\mathrm{eff}}=\frac{1}{2}$ spin-orbit insulator \SrIrOBL has been studied by means of angle-resolved photoemission spectroscopy. A comparison of the results for bilayer \SrIrOBL with available literature data for the related single-layer compound \SrIrOSL reveals qualitative similarities and similar $J_{\mathrm{eff}}=\frac{1}{2}$ band widths for both materials, but also pronounced differences in the distribution of the spectral weight. In particular, photoemission from the $J_{\mathrm{eff}}=\frac{1}{2}$ states appears to be suppressed. Yet, it is found that the \SrIrOBL data are in overall better agreement with band-structure calculations than the data for \SrIrOSL.
\end{abstract}


\maketitle

\section{Introduction}
Extending from the wealth of phenomena occurring in transition-metal oxides, which has been studied already extensively in the last century (see \eg Ref.~\onlinecite{Goodenough-ProgSolidStateChem-1971}), compounds containing heavy transition-metal elements like iridium have recently attracted much interest. Probably the most prominent example is the discovery of a $J_{\mathrm{eff}}=\frac{1}{2}$ Mott-insulating state in the material \SrIrOSL~\cite{Kim-PhysRevLett-2008}. Subsequently, various theoretical proposals ranging from the realization of the Kitaev model in Mott insulators~\cite{Jackeli-PhysRevLett-2009} to ``tailoring'' potential topological insulators out of related materials~\cite{Carter-PhysRevB-2012} have been made. In the layered quasi-two-dimensional compound \SrIrOSL it is the interplay of crystal-field splitting, spin-orbit coupling, and electron-correlation effects that determines the $5d$ electronic structure and leads to the realization of the complex $J_{\mathrm{eff}}=\frac{1}{2}$ Mott-insulating weakly ferromagnetic (canted antiferromagnetic) state~\cite{Kim-PhysRevLett-2008, Kim-Science-2009}. The very nature of this state also suggests that the low-energy electronic structure is susceptible to electron-lattice interactions (distortions, phonons) as well as changes in the magnetic structure~\cite{Moon-PhysRevB-2009}.

The corresponding Ir-O bilayer compound \SrIrOBL is expected to form a similar $J_{\mathrm{eff}}=\frac{1}{2}$ state while it is suggested to be very close to an insulator-metal transition and just on the verge of the Mott-insulating region with a largely diminished gap~\cite{Moon-PhysRevLett-2008}. While band-structure calculations indicate that the low-energy electronic structures of the two materials are overall similar~\cite{Moon-PhysRevLett-2008}, their magnetic properties somewhat differ. Magnetization measurements show a weakly ferromagnetic state in \SrIrOSL~\cite{Cao-PhysRevB-1998}; the same is true for \SrIrOBL, yet, additionally a peculiar magnetization reversal at low temperatures has been reported~\cite{Cao-PhysRevB-2002}. The true nature of the magnetic structure has been revealed in a series of resonant X-ray magnetic scattering experiments. These have established that both \SrIrOSL and \SrIrOBL actually order as commensurate antiferromagnets~\cite{Kim-Science-2009,Boseggia-PhysRevB-2012}, and in addition provide direct evidence in support of the $J_{\mathrm{eff}}=\frac{1}{2}$ model through the observation of  extremely large $L_3/L_2$ branching ratios. One very significant difference between the two compounds, however, is that in \SrIrOSL the moments are confined to the $a$-$b$ plane (with a small canting), while in  \SrIrOBL they reorient to point along the $c$ axis~\cite{Jackeli-PhysRevLett-2009,Kim-arXiv-2012, Boseggia-JPCM-2012}. It is also worth noting that neutron diffraction experiments indicate that magnetic correlations may extend to significantly higher temperatures than indicated by the resonant X-ray scattering experiments~\cite{Dhital-arXiv-2012}. Furthermore, recent resonant inelastic X-ray scattering experiments show pronounced differences in the spin excitation spectra of \SrIrOSL~\cite{Kim-PhysRevLett-2012} and \SrIrOBL~\cite{Kim-arXiv-2012-I}.

Given its entanglement with the magnetic properties it is important to provide a comparison of the low-energy electronic structure for the two compounds. A powerful experimental tool to determine the electronic spectrum is angle-resolved photoemission spectroscopy (ARPES). While ARPES data on \SrIrOSL have been reported previously~\cite{Kim-PhysRevLett-2008}, until now such data on \SrIrOBL were not available. Therefore, in this article we report the low-energy electronic structure for \SrIrOBL measured by ARPES and compare it to the available single-layer data as well as density-functional-theory band-structure calculations. We find an overall very good agreement between the experiment and the calculations for the bilayer compound, yet, the measured spectral weight from the $J_{\mathrm{eff}}=\frac{1}{2}$ states is extremely weak. Moreover, while the distribution of spectral weight is markedly different from \SrIrOSL, the data from both compounds are qualitatively similar and exhibit some disagreement with the calculations in the center of the Brillouin zone.
\begin{figure}
\centering
\includegraphics[width=0.75\columnwidth]{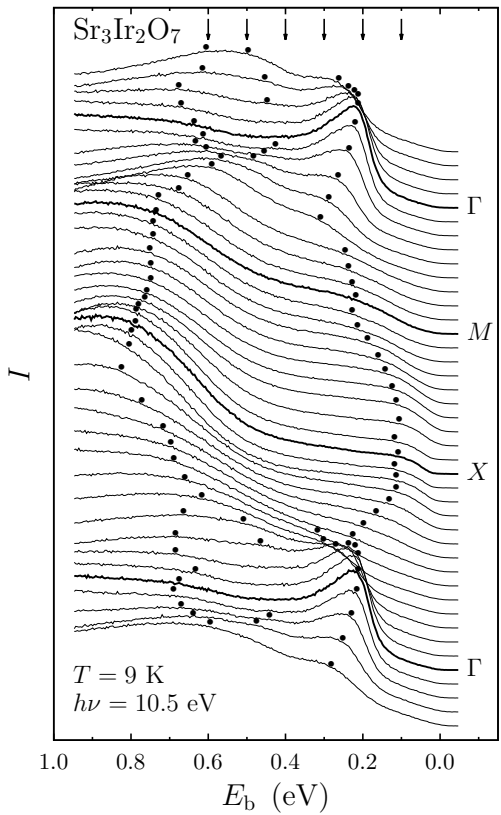}
\caption{Photoemission spectra (EDCs) at various points in the Brillouin zone along the dashed high-symmetry line in Fig.~\ref{fig:ConstantEnergyMaps}. The high-symmetry points are denoted as in Ref.~\onlinecite{Kim-PhysRevLett-2008}. The bullets ($\bullet$) indicate the local curvature maxima of the concave parts of the EDCs. The arrows at the top of the figure denote the energies for which constant-energy maps are presented in Fig.~\ref{fig:ConstantEnergyMaps}.}
\label{fig:EDCs}
\end{figure}

\section{Experimental results}
The studied single crystal of \SrIrOBL was synthesized at the Clarendon Laboratory using a self-flux technique as described in Ref.~\onlinecite{Boseggia-PhysRevB-2012}. ARPES experiments were performed at the BALTAZAR laser-ARPES facility using a laser-based angle-resolving time-of-flight analyzer and linearly polarized $10.5$~eV photons~\cite{Berntsen-RevSciInstrum-2011}. The $c$-axis-oriented single crystal of \SrIrOBL was cleaved and measured under ultra-high-vacuum conditions with a pressure below $10^{-10}$~mbar at a temperature $T=9$~K. Complementary partial areas of the Brillouin zone were covered by a series of measurements with different sample orientations and the resulting spectra were normalized and combined to yield a full data set spanning more than an eighth of the Brillouin zone. The energy and crystal-momentum resolution of the measurements are about $10$~meV and $0.02$~\AA{}$^{-1}$, respectively. The chemical potential was determined within an accuracy of about $25$~meV through a work-function estimate.
\begin{figure}[t]
\centering
\includegraphics[width=.95\columnwidth]{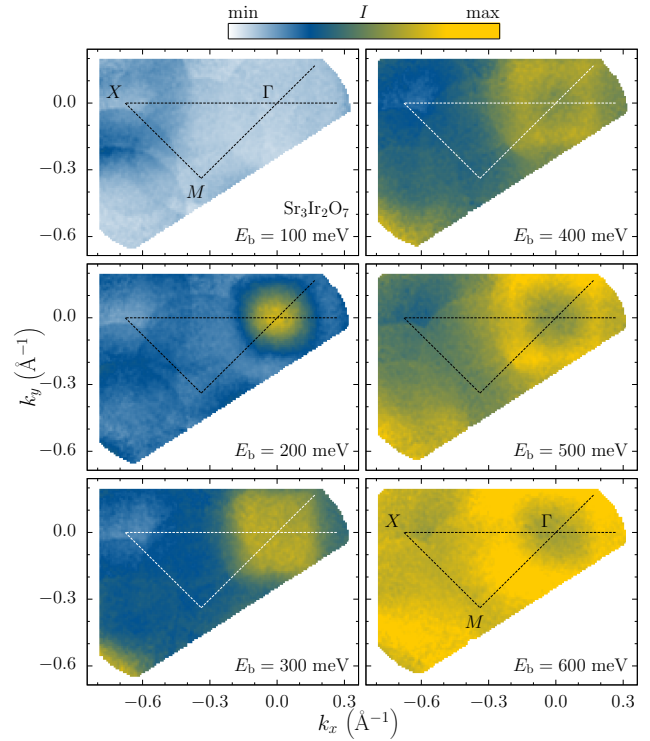}
\caption{Constant-energy surfaces at selected binding energies between $E_{\mathrm{b}}=100$~meV (\emph{top left}) and $E_{\mathrm{b}}=600$~meV (\emph{bottom right}). The dashed line indicates the high-symmetry reciprocal-space line along which spectra are presented in Figs.~\ref{fig:EDCs} and~\ref{fig:BandStructure}.}
\label{fig:ConstantEnergyMaps}
\end{figure}

Figure~\ref{fig:EDCs} shows measured ARPES spectra [energy-distribution curves (EDCs)] along a line of high-symmetry points in reciprocal space. The insulating character of \SrIrOBL is reflected in the lack of spectral weight around zero binding energy ($E_{\mathrm{b}}=0$, no Fermi surface). With increasing binding energy first a weak dispersive feature appears around the $X$ point indicated by the extremal EDC curvature. At the center of the Brillouin zone ($\Gamma$) there seems to be a very weak ``intensity tail'' at low binding energy, yet, the first clear maximum at $\Gamma$ is found just above $E_{\mathrm{b}}\approx 200$~meV. While being rather flat in the close vicinity of $\Gamma$, the spectral weight rather quickly shifts from this ``band'' to a feature at higher binding energies when departing from $\Gamma$ (in every direction) and finally forms a broad flat maximum located away from $\Gamma$ at $E_{\mathrm{b}}\gtrsim 700$~meV.

A further overview over the spectral-weight distribution in the whole Brillouin zone is provided in the constant-energy surfaces depicted in Fig.~\ref{fig:ConstantEnergyMaps}. In this context, it should be noted that the observed Brillouin zone of \SrIrOBL is smaller than expected for the originally determined tetragonal space group $I4/mmm$ with an in-plane lattice constant $a=3.896$~\AA{}~\cite{Subramanian-MatResBull-1994}. Therefore, the near-surface electronic structure provides further evidence that at least at low temperatures \SrIrOBL might feature similar coherent IrO$_6$ octahedra rotations as \SrIrOSL leading to a bigger tetragonal unit cell~\cite{Crawford-PhysRevB-1994} or even have an orthorhombic structure as reported previously (see, \eg Refs.~\onlinecite{Cao-PhysRevB-2002} or~\onlinecite{Matsuhata-JSolidStateChem-2004}).

\section{Analysis}
\begin{figure}
\centering
\includegraphics[width=.95\columnwidth]{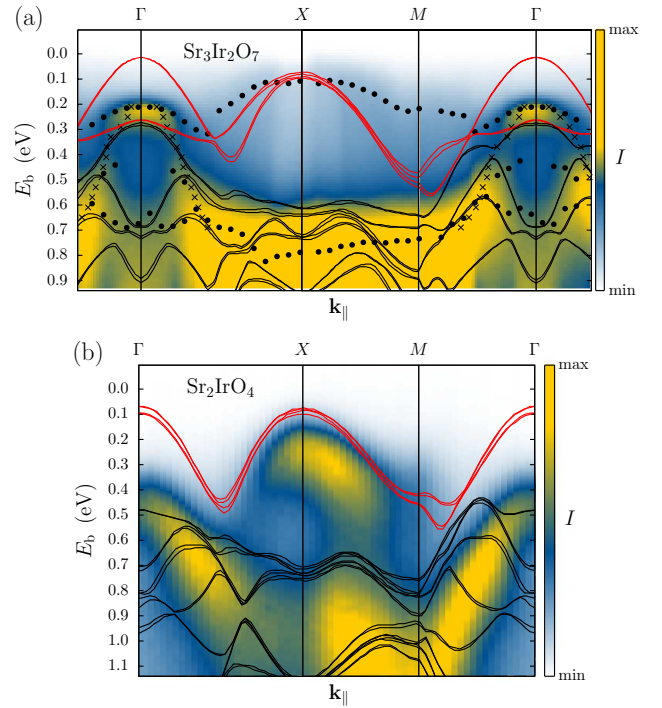}
\caption{Experimental ARPES spectra (along the dashed line in Fig.~\ref{fig:ConstantEnergyMaps}) for \SrIrOBL [$h\nu=10.5$~eV, $T=9$~K, this work, (a)] and \SrIrOSL [$h\nu=85$~eV, $T=100$~K, Ref.~\onlinecite{Kim-PhysRevLett-2008}, (b)]. The solid lines represent the calculated band structures of Ref.~\onlinecite{Moon-PhysRevLett-2008}. The red lines are $J_{\mathrm{eff}}=\frac{1}{2}$ bands, the black ones correspond to $J_{\mathrm{eff}}=\frac{3}{2}$. In (a) the bullets ($\bullet$) represent the local curvature maxima of the concave parts of the EDCs (cf. Fig.~\ref{fig:EDCs}), the crosses ($\times$) indicate MDC peaks for binding energies between $0.2$~eV and $0.7$~eV in the vicinity of $\Gamma$.}
\label{fig:BandStructure}
\end{figure}

The ARPES data on \SrIrOBL presented here are consistent with previous studies of the optical conductivity which showed an insulating state with a small gap~\cite{Moon-PhysRevLett-2008}. To facilitate a further comparison, the ARPES spectra for \SrIrOBL are depicted together with density-functional-theory band-structure calculations~\cite{Moon-PhysRevLett-2008} in Fig.~\ref{fig:BandStructure}(a). Like in Fig.~\ref{fig:EDCs}, the bullets ($\bullet$) represent distinct maxima in the concave EDC curvature. The high intensity above $700$~meV with the strong mode dispersing to about $250$~meV at $\Gamma$ show a remarkable agreement with the $J_{\mathrm{eff}}=\frac{3}{2}$ bands (solid black lines) predicted by the calculations. The distribution of the remaining spectral weight at lower energies has the same topology as the calculated $J_{\mathrm{eff}}=\frac{1}{2}$ bands (solid red lines). Hence, even though not all the bands could be clearly resolved in our experiment the data show a good overall agreement with the results of the calculations.

Turning to the pronounced feature in the vicinity of $\Gamma$ the EDCs reveal two modes originating at $E_{\mathrm{b}}\approx 250$~meV. An intense band strongly disperses towards the high-energy maximum indicated by the curvature maxima at $E_{\mathrm{b}}\gtrsim 450$~meV, a second weakly dispersing band with only low intensity follows the low-energy onset between $200$~meV and $300$~meV. Naturally, the fast dispersing feature is seen more clearly in the intensity distribution at fixed binding energies [momentum-distribution curves (MDCs)]. 
The MDCs along the high-symmetry lines close to $\Gamma$ can be well described by a model function consisting of two Lorentzians. The resulting peak positions (crosses, $\times$) are depicted in Fig.~\ref{fig:BandStructure}(a). Starting at $E_{\mathrm{b}}\approx 250$~meV they trace the intense band and overlap with the EDC curvature maxima at higher binding energies. These features identified in the EDCs and MDCs around $\Gamma$ are consistent with the calculated $J_{\mathrm{eff}}=\frac{3}{2}$ and $J_{\mathrm{eff}}=\frac{1}{2}$ bands.
Overall, we observe that all the $J_{\mathrm{eff}}=\frac{1}{2}$ bands carry very little spectral weight. While the bands around $X$ and a part of the bilayer-split bands around $\Gamma$ can still be recognized, the only sign of the predicted split-off bands at very low binding energies around $\Gamma$ might be the aforementioned almost vanishingly weak EDC ``tail'' below $E_{\mathrm{b}}\approx150$~meV (cf. Fig.~\ref{fig:EDCs}),  whose signature, however, is too faint for any further analysis.
Owing to their weakness, the width of the occupied $J_{\mathrm{eff}}=\frac{1}{2}$ bands is difficult to determine. Judging from the onset of spectral weight, given by the lowest-energy EDC curvature maxima, one obtains a band width of about $200$~meV. Yet, based on our data, it cannot be excluded that several bands are hidden in the spectral region between about $100$~meV and $600$~meV. Therefore, the most accurate estimate for a single $J_{\mathrm{eff}}=\frac{1}{2}$ band appears to be possible considering only the dispersion between $X$ and $M$ having a band width of about $120$~meV.
\section{Discussion}
Our data on \SrIrOBL also show a qualitative resemblance with the ARPES data on \SrIrOSL~\cite{Kim-PhysRevLett-2008} as there are clearly valence-band maxima at $\Gamma$ and $X$ where the latter is found at a lower binding energy. For a direct comparison the ARPES data for \SrIrOSL ($h\nu=85$~eV, $T=100$~K, Ref.~\onlinecite{Kim-PhysRevLett-2008}) and the calculated band structure of Ref.~\onlinecite{Moon-PhysRevLett-2008} are shown in Fig.~\ref{fig:BandStructure}(b). Contrary to the case of \SrIrOBL, the details of the calculations seem to differ substantially from the data particularly close to the center of the Brillouin zone and in the size of the predicted energy gap. In Ref.~\onlinecite{Kim-PhysRevLett-2008} it has been pointed out that the calculations reproduce the valence-band-maxima topology for this material. Yet, the assignment of both the valence-band maxima at $X$ and $\Gamma$ to the $J_{\mathrm{eff}}=\frac{1}{2}$ bands~\cite{Kim-PhysRevLett-2008} is ambiguous as in \SrIrOBL the main spectral weight at $\Gamma$ seems to originate from the $J_{\mathrm{eff}}=\frac{3}{2}$ states. Taking into account only the band connected to the valence-band maximum at $X$, the $J_{\mathrm{eff}}=\frac{1}{2}$ band width between $X$ and $M$ appears to be about $150$~meV which is rather close to our estimate for \SrIrOBL, while for both materials the calculated band width is about two to three times larger ($\approx 400$~meV between $X$ and $M$)~\cite{Moon-PhysRevLett-2008}. The most pronounced difference between the ARPES data of the single-layer and bilayer materials is the intensity found in the valence-band maxima---while for \SrIrOSL a high intensity is found in the valence-band maximum at $X$, the low-energy spectral weight in \SrIrOBL is substantially suppressed.

The measurements of the optical conductivity and the calculated density of states~\cite{Moon-PhysRevLett-2008} would suggest more similar results for both materials, but various effects could lead to the observed difference. In general, it could be related to varying photoemission cross-sections for the distinct photon energies used in the experiments~\cite{Yeh-AtDataNuclDataTables-1985}. However, this does not explain the weakness of the $J_{\mathrm{eff}}=\frac{1}{2}$ bands in \SrIrOBL as they derive from the same Ir $5d$ states as the $J_{\mathrm{eff}}=\frac{3}{2}$ bands. Also, there could be an effect from probing different out-of-plane crystal momenta in the not truely two-dimensional systems. Additionally, the symmetry of the $J_{\mathrm{eff}}=\frac{1}{2}$ states could play a role in the photoemission process---however, the experiment was not performed in a specific high-symmetry geometry, which renders the substantial suppression of the intensity of the $J_{\mathrm{eff}}=\frac{1}{2}$ states due to that unlikely. Eventually, the low intensity might be the result of a lack in suitable final states excitable by the rather low photon energy used in the experiment. Future ARPES experiments on both materials employing various photon energies and geometries could provide an answer to this open question.

Concerning the peculiarities in the magnetic properties of these materials, the observation of the (qualitatively) similar low-energy electronic structures in \SrIrOSL and \SrIrOBL would suggest similar magnetic properties. However, different moment orderings cannot be excluded either as long as the ``orbital order'' is similar~\cite{Kim-arXiv-2012} and might provide a reason for the marked differences observed in magnetization and muon-spin rotation~\cite{Franke-PhysRevB-2011}. Also, as mentioned before, the rather good agreement of the data and band calculations for \SrIrOBL but not \SrIrOSL could be a hint for less pronounced electron correlations in the bilayer material consistent with the apparent departure from the spin-wave expectation observed in resonant inelastic X-ray scattering experiments~\cite{Kim-arXiv-2012-I}.

\section{Conclusion}
To summarize, we presented experimental ARPES data on \SrIrOBL and compared them to data on \SrIrOSL as well as band-structure calculations available in the literature. The data on both compounds are qualitatively similar, thus confirming the related electronic structures. The comparison with the calculations shows a very good overall agreement for \SrIrOBL, thus confirming the $J_{\mathrm{eff}}=\frac{1}{2}$ insulating state in this material. However, the $J_{\mathrm{eff}}=\frac{1}{2}$ band width appears to be reduced as compared to the calculations and the observed $J_{\mathrm{eff}}=\frac{1}{2}$ bands carry very little spectral weight. Further studies are required to elucidate the origin of this effect as well as the partial disagreement of the data and calculations for \SrIrOSL.

\acknowledgments
This work was made possible through support from the Knut and Alice Wallenberg Foundation, the Swedish Research Council, the Swiss National Science Foundation, its NCCR MaNEP and Sinergia network MPBH, as well as the EPSRC.

\end{document}